\documentclass[acmlarge]{acmart}

\usepackage{booktabs} 
\graphicspath{{./images/}}
\usepackage{graphicx}
\usepackage{booktabs}
\usepackage{todonotes}
\usepackage{enumitem}

\usepackage[ruled]{algorithm2e} 

\SetAlFnt{\small}
\SetAlCapFnt{\small}
\SetAlCapNameFnt{\small}
\SetAlCapHSkip{0pt}
\IncMargin{-\parindent}

\setcopyright{rightsretained} 
\acmJournal{IMWUT}
\acmVolume{2}
\acmNumber{2}
\acmArticle{59}
\acmYear{2018}
\acmMonth{6}
\acmPrice{}\acmDOI{10.1145/3214262}

\received{August 2017}
\received[revised]{February 2018}
\received[accepted]{April 2018}

\begin{document}

\title{Discovering Smart Home Internet of Things Privacy Norms Using Contextual Integrity} 

\author{Noah Apthorpe}
\affiliation{%
  \institution{Princeton University}
  \department{Computer Science Department}
  \streetaddress{35 Olden St}
  \city{Princeton}
  \state{NJ}
  \postcode{08540}
  \country{USA}}
\email{apthorpe@cs.princeton.edu}
\author{Yan Shvartzshnaider}
\affiliation{%
  \institution{New York University}
  \department{Computer Science Department}
  \streetaddress{251 Mercer St, Room 305}
  \city{New York}
  \state{NY}
  \postcode{10012}
  \country{USA}
}
\affiliation{%
  \institution{Princeton University}
  \department{Computer Science Department}
  \streetaddress{35 Olden St}
  \city{Princeton}
  \state{NJ}
  \postcode{08540}
  \country{USA}
}
\email{yansh@nyu.edu}
\author{Arunesh Mathur}
\affiliation{%
  \institution{Princeton University}
  \department{Computer Science Department}
  \streetaddress{35 Olden St}
  \city{Princeton}
  \state{NJ}
  \postcode{08540}
  \country{USA}
}
\email{amathur@cs.princeton.edu}
\author{Dillon Reisman}
\affiliation{%
  \institution{Princeton University}
  \department{Computer Science Department}
  \streetaddress{35 Olden St}
  \city{Princeton}
  \state{NJ}
  \postcode{08540}
  \country{USA}
}
\email{dreisman@princeton.edu}
\author{Nick Feamster}
\affiliation{%
  \institution{Princeton University}
  \department{Computer Science Department}
  \streetaddress{35 Olden St}
  \city{Princeton}
  \state{NJ}
  \postcode{08540}
  \country{USA}
}
\email{feamster@cs.princeton.edu}


\begin{abstract}
The proliferation of Internet of Things (IoT) devices for consumer ``smart'' homes
raises concerns about user privacy.
We present a survey method based on the Contextual Integrity (CI) privacy framework
that can quickly and efficiently discover privacy norms at scale.
We apply the method to discover privacy norms in the smart home context, surveying
1,731 American adults on Amazon Mechanical Turk.
For \$2,800 and in less than six hours, 
we measured the acceptability of 3,840 information flows
representing a combinatorial space of smart home devices sending consumer information to first and third-party recipients under various conditions. 
Our results provide actionable recommendations for IoT device manufacturers, including design best practices and instructions for adopting our method for further research.
\end{abstract}

%
%

 
%
%


\maketitle

\renewcommand{\shortauthors}{N. Apthorpe et al.}

\section{Introduction}
\label{sec:intro}
Internet of Things (IoT) devices for consumer ``smart'' homes can observe sensitive
details about users' in-home activities and transmit this information on the Internet~\cite{iot-exercise, motherboard-teddy, iot-medical, iot-sex}. 
The sensitivity of smart home data and the nascency of the IoT calls for
effective scalable methods to discover and understand people's privacy norms regarding these devices.
Understanding these privacy norms will allow manufacturers to design devices that
consumers are comfortable incorporating into their homes and help government regulators and consumer advocates identify and contextualize privacy violations.
Several previous studies have attempted to measure privacy norms pertaining to IoT
devices (Section~\ref{sec:related}) but 
none discover privacy norms at scale in a manner that is based on a formal theory
of privacy.

In this paper, 
we present a general, scalable survey method for discovering consumer privacy norms
based on the
\textit{Contextual Integrity (CI)} privacy framework~\cite{nissenbaum2010privacy} (Section~\ref{sec:methods}). 
CI is a well-established theory that defines privacy norms as the generally accepted
appropriateness of specific information exchanges, or ``information flows,'' in
specific contexts. Information flows and associated contexts can be described using five parameters: sender, recipient, subject, attribute, and transmission principle.
This precise formulation makes it possible to thoroughly investigate the combinatorial
space of contextual information flows and associated privacy norms
with an automated, large-scale survey on a crowdsourcing platform.
Our use of CI 
also ensures that the method is repeatable,
both for the same types of devices over time, as well as for entirely new classes
of devices.
  
The method we develop is effective for discovering privacy norms
in general. In this paper, we focus on applying the method to discover smart home
privacy norms.
We conducted a survey with a population of 1,731 adults from the United States on the Amazon Mechanical Turk (MTurk) platform. 
The survey cost \$2,800 and allowed us to query the acceptability of 3,840 information flows involving smart home devices in less than six hours and identify associated privacy norms (Section~\ref{sec:results}).
Our results provide insightful observations and actionable recommendations
for IoT device manufacturers, regulators, and consumer advocates (Section~\ref{sec:discussion}).

Device manufacturers can use our survey method to perform their own research
on how consumers might view the use of data that their products collect.
We designed the method to make it easy to customize
with new information flows and contexts, allowing manufacturers to discover privacy
norms relevant to specific products, including ones we have not studied in this
paper.
The results will indicate whether existing or proposed devices may violate
established privacy norms, providing an opportunity to preempt negative user feedback,
public relations debacles, or regulatory scrutiny.
The method is also relatively inexpensive, allowing manufacturers to investigate
a broad set of privacy norms for a fraction of product development costs or conventional
surveys.

Prevailing trends and patterns in our results suggest several best practices for
smart home device manufacturers.
Manufacturers should note that information flows not specifically related
to core device features are universally viewed as egregious violations of
privacy norms.
For example, the results quantifiably demonstrate that a fitness tracker sending
recorded audio is considerably less acceptable than the same device sending exercise data.
Information flows to Internet service providers (ISPs), for advertising, or for
indefinite storage were also especially unacceptable across all devices.
In general, many types of information flows were viewed unfavorably by consumers,
suggesting that setting default limits on the types of data collected and where that data
is sent is a reasonable starting point for meeting consumer privacy expectations.

Regulators with rulemaking authority (e.g., the United States Federal Communication
Commission, which has previously written rules on privacy \cite{fccResolution}) could use our method
to discover existing privacy norms and to better inform rulemaking.
For example, we find that despite increased cultural awareness of the ``smart home,'' the average survey respondent still views information flows from smart home devices to recipients outside of the home as generally unacceptable unless the device owner has specifically granted consent. In light of known issues with current consent mechanisms (privacy policies and EULAs)~\cite{mcdonald2008cost}, this result further motivates the implementation of rules
requiring granular consent options within clearly specified contexts~\cite{martin2016measuring}.

Consumer advocacy groups could use our method to determine
what device behaviors are of particular concern. This could help direct efforts to pressure manufacturers into changing practices.
For example, our results found that social media accounts were one of the most unacceptable recipients of user information from smart home devices. This finding might motivate consumer advocacy
groups to create guidelines for how smart home devices might better integrate data
sharing capabilities with social media platforms.

In summary, this work makes the following contributions:
\begin{enumerate}
\item We present a survey method for privacy norm discovery that integrates a formal theory of privacy with combinatorial testing at scale.
\item We provide insights into existing privacy norms in the smart home context, contributing to the ongoing discussion about the evolving privacy impact of IoT technology \cite{asseo2016internet, daugherty2015driving, fcc-greenpaper}.
\item We provide actionable recommendations based on discovered privacy norms, including development best practices for device manufacturers  and instructions for further research using our survey method.
\end{enumerate}
\noindent
We encourage others---including manufacturers, regulators, consumer advocates, and academics---to
use our CI survey method for their own research. Follow-up studies could focus on
different aspects of IoT devices, as well as technical innovation in other
contexts (e.g., mobile application development). Repeated surveys using our method
could also provide data for longitudinal analyses of how cultural privacy norms
change in response to rapidly evolving technology. 

\section{Contextual Integrity Background}
\label{sec:background}
The rapid pace of technological innovation introduces new ways to share and acquire information, forcing us to rethink established definitions of privacy. Conventional privacy definitions, such as Role-base Access Control (RBAC)~\cite{ferraiolo2003role,ferraiolo2001proposed}, Attributed Base Access Control (ABAC)~\cite{parducci2005extensible}, and Enterprise Privacy Authentication Language (EPAL)~\cite{ashley2003enterprise,EPAL-website}, focus on classifying information types and establishing sophisticated access control mechanisms to prevent information from reaching unintended recipients. They do not incorporate the many contexts and norms that govern information exchange. In our daily activities, we traverse many environmental settings and take on many different roles. For example, many people work from home or during transit, often communicating personal information from work and work information from home. 

CI defines privacy in terms of conformance of a given informational exchange to contextual information norms~\cite{nissenbaum2010privacy}. Norms come as a part of a context, or a specific setting established by law, policy, common practice, social pressures, and beliefs. An information flow that is misaligned with established norms potentially violates privacy expectations in the given context. CI is becoming more relevant as the lines between professional, personal, and public spaces are shifted and blurred by the increasingly rapid pace of information exchange.

CI describes norms in terms of five information flow parameters: (1)~the \emph{sender} of the information, (2)~the \emph{recipient} of the information, (3)~the \emph{attribute} or type of information, (4)~the \emph{subject} of the information, and (5)~a \emph{transmission principle} that states the condition under which the information flow is permitted. A change to any parameter may cause a privacy norm violation. For example, you (the sender/subject) might feel that it is acceptable to provide your doctor (recipient) with medical records (attribute) with the requirement of confidentiality (transmission principle). However, if the recipient is your boss, you might reconsider. 

CI has been used in many legal and computer science studies as an alternative framework to define and reason about privacy
\cite{shvartzshnaider2016learning,grodzinsky2010applying, shi2013using}. 
The computer science community has focused on formal expression of contextual privacy norms to detect infraction using formal logic and to propose methods for accountability and enforcement~\cite{chowdhury2013privacy, barth2006privacy,criado2015implicit}.
Legal privacy scholars have used CI as a new lens to re-evaluate existing regulations, such as the meaning of ``reasonable expectation of privacy'' in the Fourth Amendment ~\cite{selbst2013contextual}, and to find gaps in the privacy settings of existing systems, including Facebook~\cite{hull2011contextual} and Google~\cite{zimmer2008privacy}.
As a common framework, CI has inadvertently fostered synergy between these communities, helping them advance the common goal of preserving technology users' privacy. 

In addition to providing a way to capture and evaluate information flows against established privacy norms, CI can also capture new and evolving norms, such as when new technology is introduced. In this case, CI argues for reassessing information flow parameters with respect to ``their merits as a function of their meaning and significance in relation to the aims, purposes, and values of the context''~\cite{nissenbaum2010privacy}. When toasters begin transmitting information to the cloud, they become senders in an entirely new set of information flows and associated privacy norms. 

\section{Survey Method}
\label{sec:methods}
This section describes our method of discovering privacy norms by surveying information flow acceptability.
We chose information flows relevant to smart home devices, but the method could be used to discover privacy norms in any context.

\subsection{Selecting CI Information Flow Parameters}
\label{sec:element-selection}
The first step in our survey method is to select CI information flow parameters (senders, recipients, attributes, subjects, and transmission principles) relevant to the context of interest. 
We chose parameters relevant to smart homes by  
surveying the academic literature and popular press coverage about consumer IoT devices (Table~\ref{fig:tuple-elements}). 
These lists are not exhaustive, 
but they cover a range of smart home information flows and demonstrate the generality of our method.
Device manufacturers, regulators, consumer advocates, or academic researchers could use our survey method with different parameters in order to discover privacy norms about specific devices or information collection practices.

\begin{table}[t]
\centering
\footnotesize
\caption{CI parameter values chosen to generate smart home information flows. The \textit{subject} parameter is not listed and was set to ``its owner", referring to the owner and primary user of a device. We included a ``null'' transmission principle to generate unconditional information flows.}
\begin{tabular}{lllllll}
\textbf{Sender} & \textbf{Recipient} & \textbf{Attribute} & \textbf{Transmission Principle}\\
\toprule
a sleep monitor & the local police & \{subject\}'s location & if \{subject\} has given consent \\
a security camera & government intelligence agencies & \{subject\}'s eating habits & if \{subject\} is notified\\
a door lock & \{subject\}'s doctor & the times \{subject\} is home & if the information is kept confidential \\
a thermostat & an Internet service provider & \{subject\}'s exercise routine & if the information is anonymous\\
a fitness tracker & its manufacturer & \{subject\}'s sleeping habits & if the information is used to perform \\ 
a refrigerator & other devices in the home & audio of \{subject\} & \quad maintenance on the device\\
a power meter & \{subject\}'s immediate family & video of \{subject\} & if the information is used to provide\\
a personal assistant & \{subject\}'s social media accounts & \{subject\}'s heart rate & \quad a price discount\\
\quad (e.g. Amazon Echo) & & the times it is used & if the information is used for advertising\\
& & & if the information is used to develop\\
& & & \quad new features for the device\\
& & & if the information is not stored\\
& & & if the information is stored indefinitely\\
& & & if its privacy policy permits it\\
& & & in an emergency situation\\
& & & \textit{null (no transmission principle)}\\
\midrule
\end{tabular}
\label{fig:tuple-elements}
\end{table}

\subsubsection{Sender}
We chose the list of senders as a variety of commercially available smart home IoT devices. 
The devices represent a range of potential privacy concerns, including physical presence (security camera and door lock), various behaviors (sleep monitor, refrigerator, and fitness tracker), and energy usage (power meter and thermostat). 
This list also includes device types that many consumers would have heard about in the popular press, including ``a [smart] thermostat'' (e.g., a Nest thermostat).

We chose not to provide specific device names, such as ``Sense Sleep Monitor,'' to limit the effect of participants' opinions about specific companies on survey responses. 
``A personal assistant (e.g., Amazon Echo)'' is an exception, because ``a personal assistant'' requires additional explanation to ensure participants envision the correct type of device.  

\subsubsection{Recipient.}
We chose the list of recipients to capture the range of first and third-parties that can obtain user information from existing smart home devices. 

We included device manufacturers because they receive the most consumer information from IoT devices. Many devices communicate sensor recordings and user interactions directly to manufacturer-operated cloud servers. Device manufacturers usually provide privacy policies and enter end-user license agreements with consumers; however, these policies many not reflect generally accepted privacy norms. 

We included the local police and government intelligence agencies in consideration of recent court cases involving data obtained from IoT and mobile devices \cite{echo-trial}. 

We included ISPs because of recent scrutiny over ISP access to potentially private consumer data. The Federal Communications Commission's (FCC) Broadband Privacy Rules, passed in October 2016 and overturned by Congress in April 2017 \cite{fcc-pr}, were intended to limit the ability of ISPs to collect and sell consumer information obtained by network monitoring.  Existing academic literature also demonstrates that ISPs can infer user behaviors from IoT network traffic even when the traffic is encrypted \cite{apthorpeIoT}.

We included other devices in the home because  
several large manufacturers have IoT ecosystems that cooperate to control multiple devices in a smart home. These ecosystems include Apple HomeKit, Samsung SmartThings, and the many devices that can be controlled via an Amazon Echo or Google Home. Communications between IoT devices raise additional privacy concerns, because devices may be made by different manufacturers (with different privacy policies) or may allow sensitive information to be inferred through the analysis of data from multiple devices. 

We included the immediate family of a device owner in order to investigate the privacy norms of IoT devices acting as intermediaries in information flows otherwise acceptable in non-IoT contexts. 
A user's immediate family typically knows the user's sleeping habits and when the user is home. We were curious whether being able to learn this information about an immediate family member via an IoT device violates privacy norms. 

We included social media accounts because some available IoT devices have the ability to post directly to user profiles (after user setup and password entry) \cite{snap-spectacles}. We expect that more manufacturers will connect IoT devices to social networks in the future.

\subsubsection{Attribute.}
We chose the list of attributes to incorporate a variety of information types that a recipient could obtain about a IoT device owner from device communications. Some of these attributes are raw sensor recordings (e.g., audio, video, location) with obvious privacy implications.  Others are user behaviors that can be inferred from sensor recordings by the IoT device itself or through post-hoc analysis by the recipient (e.g., eating habits, sleeping habits, when the user is home, etc.).

\subsubsection{Subject.}
We chose the IoT device owner as the only subject in our survey. This helped limit the number of questions and reduced cognitive fatigue for participants. ``Owner'' is less jargon than ``user,'' and an IoT device's owner is the must likely subject of sensor and interaction information recorded by the device. 

We also considered several additional subjects that could be the focus of follow-up studies.  For example, the device owner's child, spouse, roommate, or guest are all individuals who could interact with IoT devices in a smart home and be the subject of information flows with different privacy norms. 

\subsubsection{Transmission Principle.}
The transmission principle parameter provides a condition under which an information flow occurs. Our list of transmission principles cover a variety of use cases for information from IoT devices. Some of these cases are specifically mentioned in many device privacy policies, such as ``if the information is used to develop new features for the device'' or ``if the information is kept anonymous.'' Other transmission principles involve the storage of the information (``indefinitely'' and ``not stored'') or previous actions by the subject (``if \{subject\} has given consent'' and ``if its privacy policy permits it'').  We also included a \textit{null} entry in our list of transmission principles as a control to measure the acceptability of unconditional information flows.  

Variations in survey responses with varying transmission principles indicate that IoT privacy norms depend on how information will be used and under what circumstances it will be transmitted. 
Information about these conditional norm variations is important for IoT device manufacturers to avoid violating consumer trust and for policymakers to regulate the IoT in ways that align with consumer expectations.  

\subsection{Survey Design}
The next step in our method is to create a survey that queries the acceptability of information flows generated from selected CI parameters.
The following sections describe each part of our survey in detail. 

\subsubsection{Consent \& Survey Overview}
\begin{figure}[t]
\includegraphics[width=0.83\textwidth]{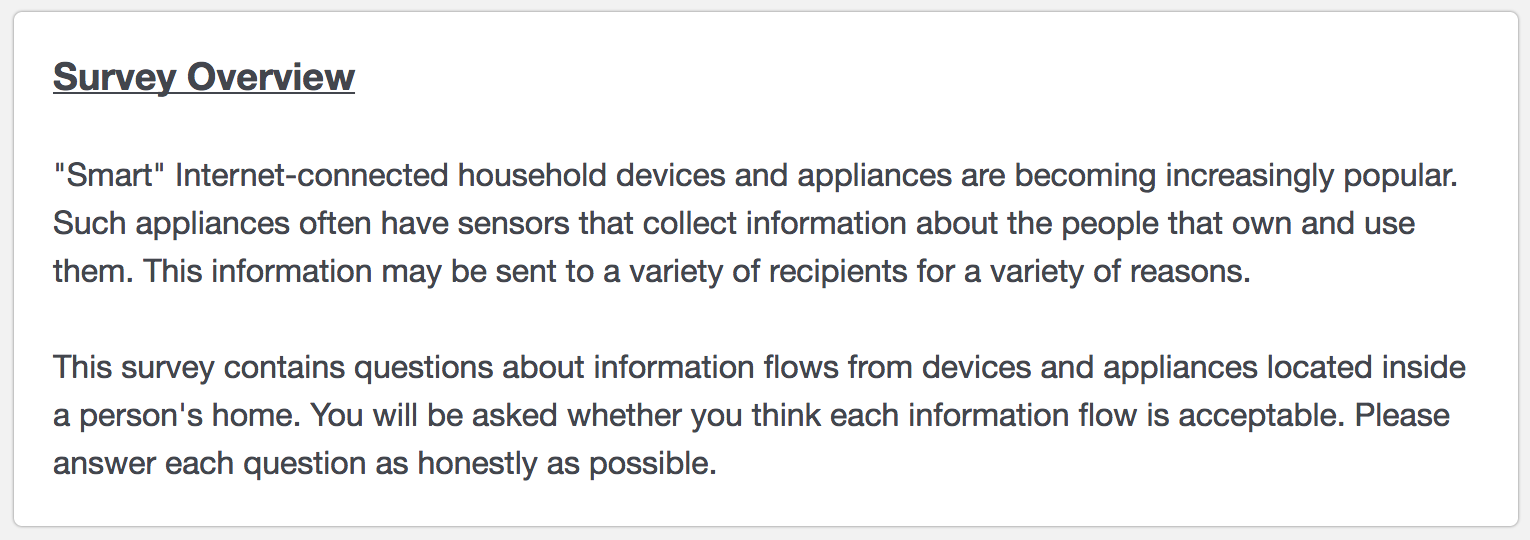}
\caption{Overview presented to participants at the beginning of the survey. The overview provides context and instructions without priming participants to view information flows as inherently acceptable or unacceptable.}
\label{fig:survey-overview}
\end{figure}

The first page of the survey contained a consent form. Participants who did not consent were prevented from taking the survey. The following page provided a brief background on the Internet of Things and an introduction to the CI questions (Figure~\ref{fig:survey-overview}).

\subsubsection{Contextual Integrity Questions} 
\label{sec:methods-ci-questions}

The main section of the survey contained questions about the acceptability of information flows generated from our lists of CI parameters (Table~\ref{fig:tuple-elements}). 
We considered all possible combinations of the parameter lists and eliminated combinations that did not make sense given the current generation of IoT devices. For example, we ruled out all combinations with a refrigerator sending its owner's heart rate to any receiver under any transmission principle. While it is conceivable that an IoT refrigerator could be designed to measure a user's heart rate, we are unaware of any existing refrigerators with heart rate monitors. Filtering such less intuitive combinations allowed us to incorporate more parameter values into the survey while limiting the total number of questions. We ended up with  3,840 five-parameter information flows after filtering. 

We next generated questions querying the acceptability of all 3,840 five-parameter information flows. 3,840~questions is obviously too many to ask an individual participant, so we divided the flows into 48 sets, where all information flows in a set had the same sender and attribute. Each participant was randomly assigned a set and asked questions about all flows in that set. For example, an individual participant might have been asked questions about a sleep monitor sending audio of its owner to various recipients under various transmission principles.

In addition to reducing the number of questions asked to each participant, grouping the information flows into sets by sender and attribute assured the independence of responses across receivers and transmission principles. By preventing participants from taking the survey more than once, we are certain that all participants answered questions about all transmission principles and all recipients. This made it possible for us to conduct significance testing across recipients and transmission principles.

\begin{figure}[t]
\centering
\includegraphics[width=0.495\textwidth]{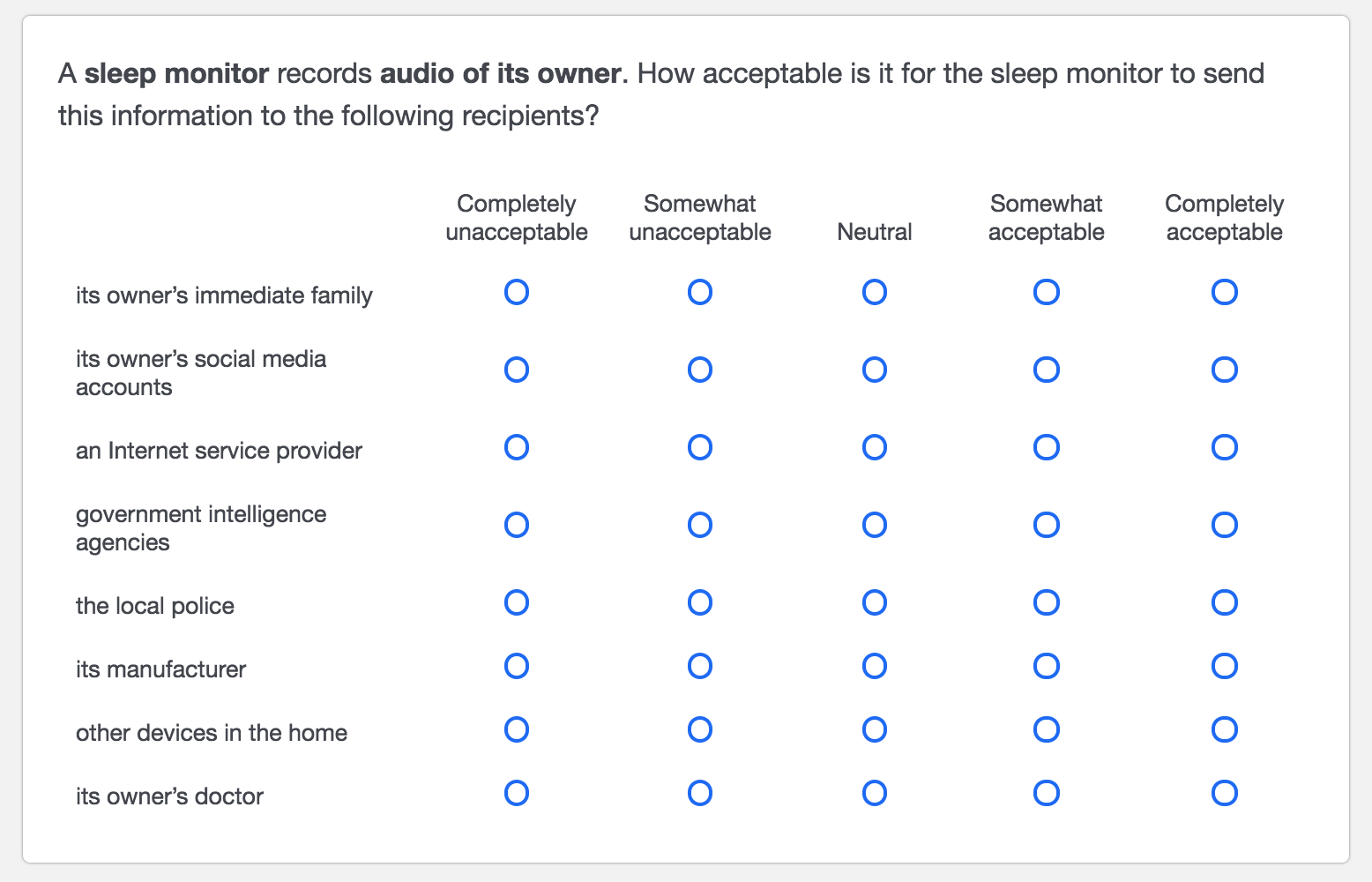}
\includegraphics[width=0.495\textwidth]{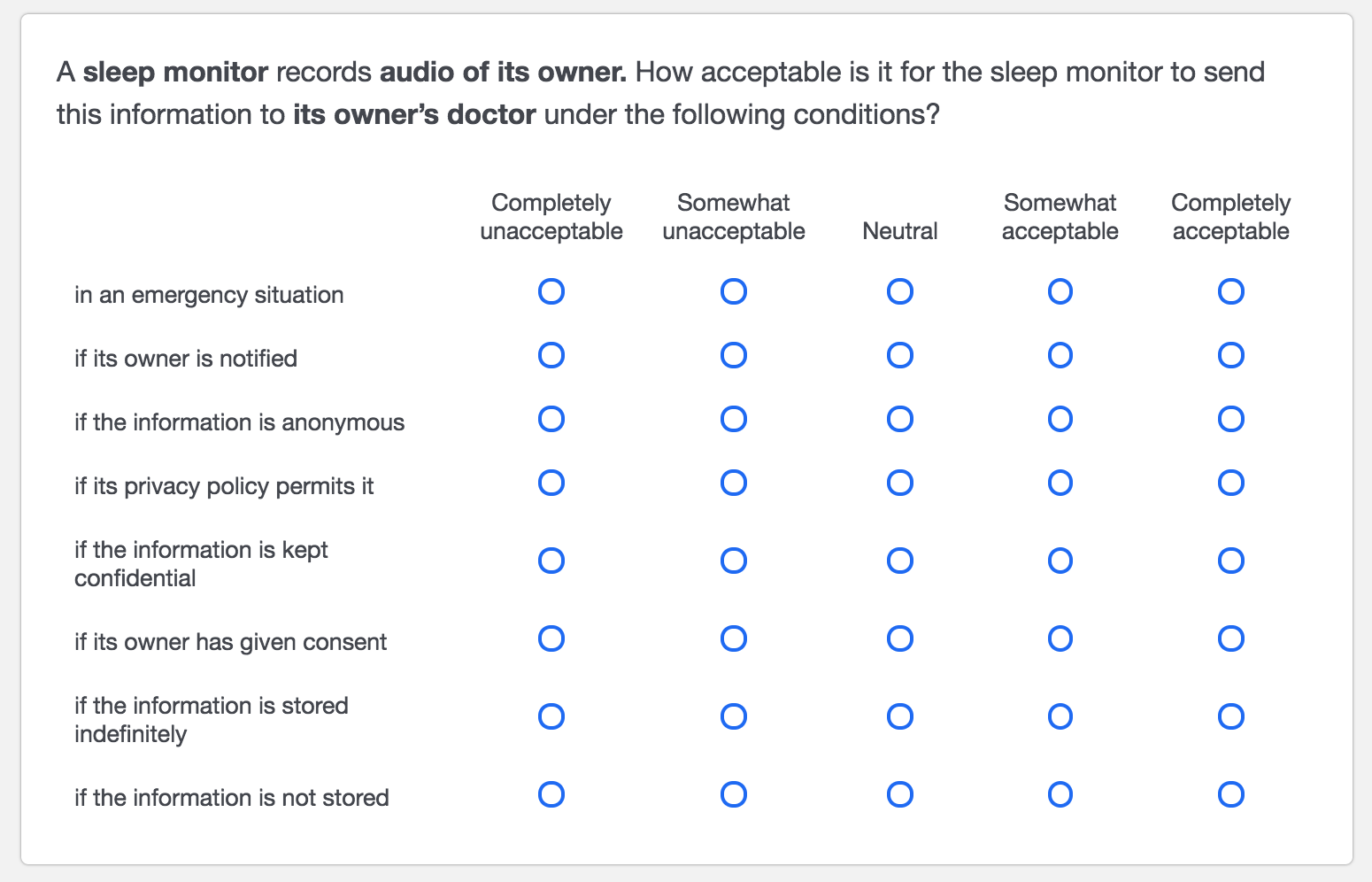}
\caption{Example matrix questions presented to survey respondents. Each respondent first saw one matrix question about unconditional information flows (\textit{null} transmission principle) with varying recipients (\textit{left}). The respondent then saw a series of questions for each recipient about information flows with varying transmission principles (\textit{right}).}
\label{fig:matrix-example}
\end{figure}

To further reduce participant cognitive load, we combined sets of multiple information flows into a question matrix format (Figure~\ref{fig:matrix-example}). The columns of each matrix were a five-point acceptability scale: ``Completely Unacceptable'', ``Somewhat Unacceptable'', ``Neutral'', ``Somewhat Acceptable'', and ``Completely Acceptable.'' The rows of each matrix contained the parameter values that varied across the flows in the matrix. We randomized the order of the rows across matrices and participants. The non-varying parameters in each matrix were set in bold font. Each participant rated 82 total information flows over 9 question matrices.

The first question matrix presented to each participant corresponded to information flows with varying receivers and the \textit{null} transmission principle (i.e., always transmit). We always initiated the survey with this question to prevent priming effects of the other transmission principles.
The remaining question matrices presented to each participant corresponded to information flows with a fixed recipient and varying transmission principles. The order of these matrices was randomized for each participant. Figure~\ref{fig:matrix-example} shows an example of both question matrix formats. 
We also inserted an attention check question (``Select `Somewhat Acceptable'\,\!'') as one row of one question matrix on each participant's survey.  
 
\subsubsection{Demographics and Technical Background Questions} 
The final section of the survey had a series of demographics and technical background questions presented to every participant, followed by a field for optional open-ended comments.
The Appendix details the self-reported demographics of survey participants.
The participants were split equally between males and females, on terms with the general US Internet user population~\cite{wnow}. A large fraction of the participants had a Bachelor's degree or higher and were also younger compared to the US population~\cite{wnow}. These differences are largely due to the MTurk platform, which tends to attract more technology-savvy users \cite{mturkeducated}.
Further, 36\% of survey respondents report owning at least one smart home device (``a `smart' (Internet-connected) device or appliance in [their] home besides a smartphone, tablet, laptop, or desktop computer''). 78\% of the respondents who reported owning at least one smart home device set up the device(s) themselves.
 
\subsection{Survey Deployment}
We created and hosted our survey using the Qualtrics~\cite{qualtrics} platform. The survey content and participant recruitment process were approved by our institution's Institutional Review Board (IRB).
We tested the survey using UserBob~\cite{userbob-website}, a usability-testing service that recruits MTurk workers to record their screen while interacting with a website and providing audio commentary about their experience. We collected five 8--15 minute interactions with our survey from UserBob. All five UserBob workers 
completed the survey within 15 minutes (even while recording audio feedback). We used their feedback and misunderstandings to subsequently refine question wordings. We did not include responses from the UserBob workers in the final results. This practice, called ``cognitive interviews'' \cite{sudman1996thinking}, is common in survey design and development.

We administered the final version of the survey using TurkPrime~\cite{litman2017turkprime}, an online service for researchers to easily create and manage MTurk Human Intelligence Tasks (HITs). We recruited 2,000 total participants and paid each participant \$1.00 for completing the survey. Running the entire survey cost \$2,800 and took less than six hours to complete. The survey was limited to workers in the US with a 90--100\% HIT approval rating.  We did not impose any limit on the prior number of  approved HITs.  Workers were automatically paid upon entering a completion code displayed at the end of the survey. Collecting all 2,000 responses took less than six hours. To avoid priming workers during task selection, we advertised the survey as a ``Home Technology Survey,'' not as a survey specifically about privacy.

\subsection{Response Analysis}
\label{sec:analysis-method}
The results from our survey method are amenable to a variety of analysis techniques. We employed the following straightforward aggregation approach to discover privacy norms pertaining to specific CI parameters. Adopters of our method could apply other analysis techniques, including multivariable modeling, to explore more complex relationships in the survey response data.

\subsubsection{Filtering \& Scaling Responses}
We first filtered the responses to remove those from 269 participants who misanswered the attention check question. This yielded a total of  1731 responses with an average of 37 responses per CI matrix question (standard deviation 1.3).  
We used a Likert scale for the responses with the following Likert items: Completely Acceptable ($2$), Somewhat Acceptable ($1$), Neutral ($0$),  Somewhat Unacceptable~($-1$),  Completely Unacceptable~($-2$). 

\subsubsection{Average Acceptability Scores}
We discover privacy norms by analyzing how different combinations of CI parameters affect survey responses. 
We grouped all information flows with the same sender and attribute and all information flows with the same transmission principle and recipient. 
These sender/attribute and recipient/transmission principle groupings coordinated with how the questions were presented to participants (Section~\ref{sec:methods-ci-questions}).
Each recipient/transmission principle group contains flows with all tested pairs of senders and attributes, while each sender/attribute group contains flows with all tested pairs of recipients and transmission principles. 
Every participant scored exactly one information flow in every recipient/transmission principle group, while every participant scored all flows in exactly one sender/attribute group.

We then computed the average\footnote{There was little difference between average and median acceptability scores due to the limited dynamic range of the Likert scale.} Likert acceptability score of all flows in each group. 
This statistic allows us to compare the pairwise effects of CI parameters on information flow acceptability.
If a parameter has a notably high or low average acceptability score across all pairwise groups (e.g., a sender with a high average acceptability score regardless of attribute), this implies that the parameter is individually important to participants' privacy norms. 
If a pair of parameters has a notably high or low average acceptability score, this implies that there is likely a privacy norm involving the interplay between the two parameters. 

\subsubsection{Significance Test}
\label{sec:methods-stats}
The division of CI survey questions into sets as described above ensures that responses across pairs of senders and attributes are independent. This allows us to perform non-parametric Wilcoxon signed-rank tests \cite{lowry2014concepts} to measure the effect of recipients and transmission principles. 

To study the effect of transmission principle, we 
perform the Wilcoxon test to compare acceptability scores between pairs of informations flows with the same sender, attribute, and recipient, but with \textit{null} and non-\textit{null} transmission principles.
For example, we compare scores of the flow ``A fitness tracker records its owner's exercise routine and sends this information to its owner's doctor'' against scores of the flow  ``A fitness tracker records its owner's exercise routine and sends this information to its owner's doctor \textit{if its owner has given consent}.''

To study the effect of recipient on information flow acceptability, we chose the recipient ``its owner's immediate family'' as a baseline. We perform the Wilcoxon test to compare acceptability scores of flow pairs with baseline and non-baseline recipients and the same other parameters. 
For example, we compare scores of the flow ``A refrigerator records its owner's eating habits and sends this information to its owner's immediate family if its owner is notified'' against scores of the flow  ``A refrigerator records its owner's eating habits and sends this information to \textit{its manufacturer} if its owner is notified.''

Since we conducted several such tests---one for each recipient and transmission principle---we accounted for multiple comparisons using the Bonferroni correction method \cite{MPT}. We took the standard 0.05 cutoff threshold and divided it by the number of tests to generate the new threshold. For transmission principle, we use a threshold of $0.00008$ ($0.05$ / $576$). For recipient, we use a threshold of $0.0001$ ($0.05 / 336$).

\section{Results}
\label{sec:results}
Our analysis of the survey responses provides insight into privacy norms in the smart home context.  In this section, we describe the results of the analysis procedures described in Section~\ref{sec:analysis-method}.  

\subsection{Average Acceptability Scores}
\begin{figure}[tp]
\includegraphics[width=0.87\textwidth]{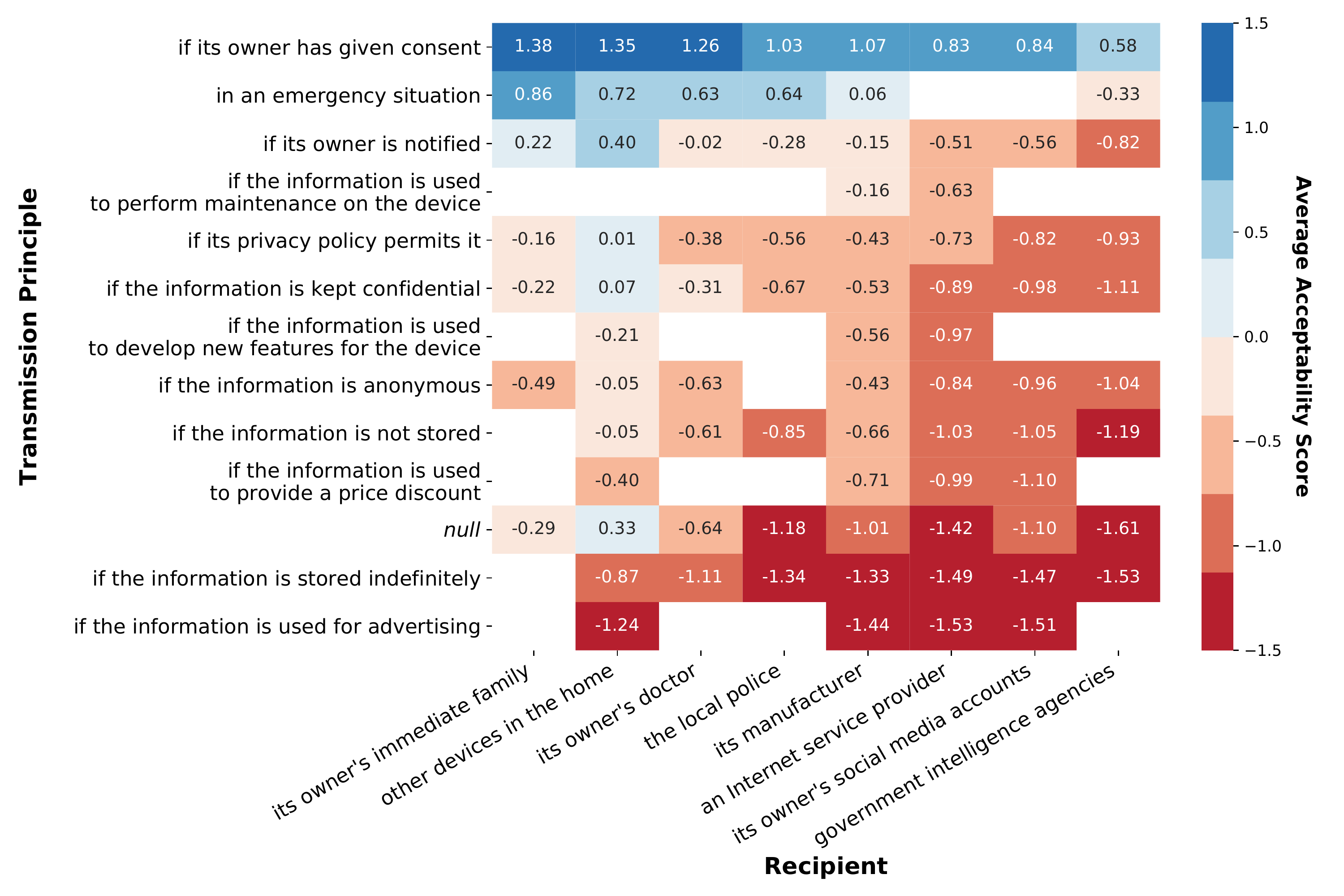}
\includegraphics[width=0.87\textwidth]{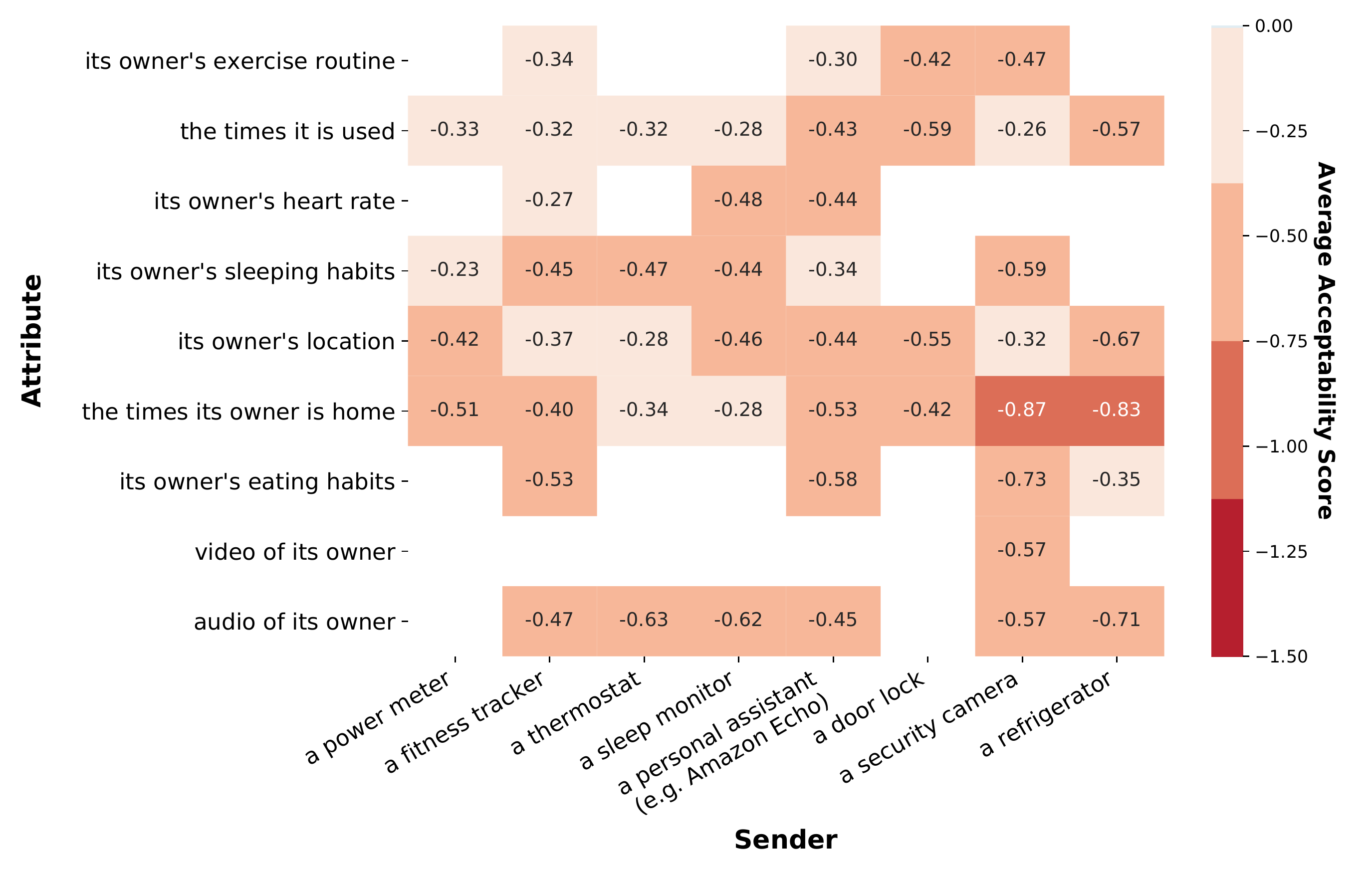}
\vspace{-10pt}
\caption{Average acceptability scores of information flows with given recipient/transmission principle pairs (\textit{top}) and given sender/attribute pairs (\textit{bottom}). 
Empty locations correspond to less intuitive information flows that were excluded in the survey. 
Parameters are sorted from top to bottom and left to right by descending average acceptability score for \textit{all} information flows containing that parameter.}
\label{fig:heatmaps}
\end{figure}

For each sender/attribute and recipient/transmission principle pair, we calculate the average acceptability score of all information flows containing that pair of parameters (Figure~\ref{fig:heatmaps}).
The only parameters with positive average acceptability scores in any pair are the transmission principles ``if its owner has given consent'' (scores range from $0.58$ to $1.38$), ``in an emergency situation'' ($-0.33$ to $0.86$), and ``if its owner is notified'' ($-0.82$ to $0.40$) and the recipients ``its owner's immediate family'' ($0.49$ to $1.38$) and ``other devices in the home'' ($-1.24$ to $1.35$). The transmission principle ``if its owner has given consent'' is notable as the only parameter with positive pairwise acceptability scores regardless of recipient.

The parameters with the lowest pairwise average acceptability scores are the transmission principles ``if the information is used for advertising'' ($-1.51$ to $-1.24$) and ``if the information is stored indefinitely'' ($-1.53$ to $-0.87$) and the recipients ``government intelligence agencies'' ($-1.52$ to $0.58$), ``its owners social media accounts'' ($-1.51$ to $0.84$) and ``an Internet service provider'' ($-1.53$ to $0.83$). Additionally, ``if the information is used for advertising'' and ``if the information is stored indefinitely'' are the only transmission principles with average acceptability scores across all flows containing those parameters below that of the same flows with the \textit{null} transmission principle.

There is less variation in average acceptability score across sender/attribute pairs ($-0.23$ to $-0.87$) than across recipient/transmission principle pairs ($-1.53$ to $1.38$). 
The senders ``a power meter'' ($-0.51$ to $-0.23$) and ``a fitness tracker'' ($-0.53$ to $-0.27$) are the most acceptable across attributes, while ``a security camera'' ($-0.87$ to $-0.26$) and ``a refrigerator'' ($-0.87$ to $-0.26$) are the least.

Interactions between CI parameters also have noticeable effects on acceptability scores. 
For example, the recipient ``other devices in the home'' has positive average acceptability scores when paired with six transmission principles but negative scores when paired with five other transmission principles.
The transmission principle ``in an emergency situation'' has positive average acceptability scores when paired with all recipients except ``government intelligence agencies.''
The attribute ``its owner's eating habits'' has a less negative score when paired with the sender ``a refrigerator'' ($-0.35$) than its scores with all other senders ($-0.73$ to $-0.53$). 
These cases highlight that CI parameters are not independent, but combine to to create meaningful contexts relevant to privacy norms. 

\subsection{Effect of Transmission Principle and Recipient on Information Flow Acceptability}
\begin{figure}[tp]
\centering
\includegraphics[width=0.85\textwidth]{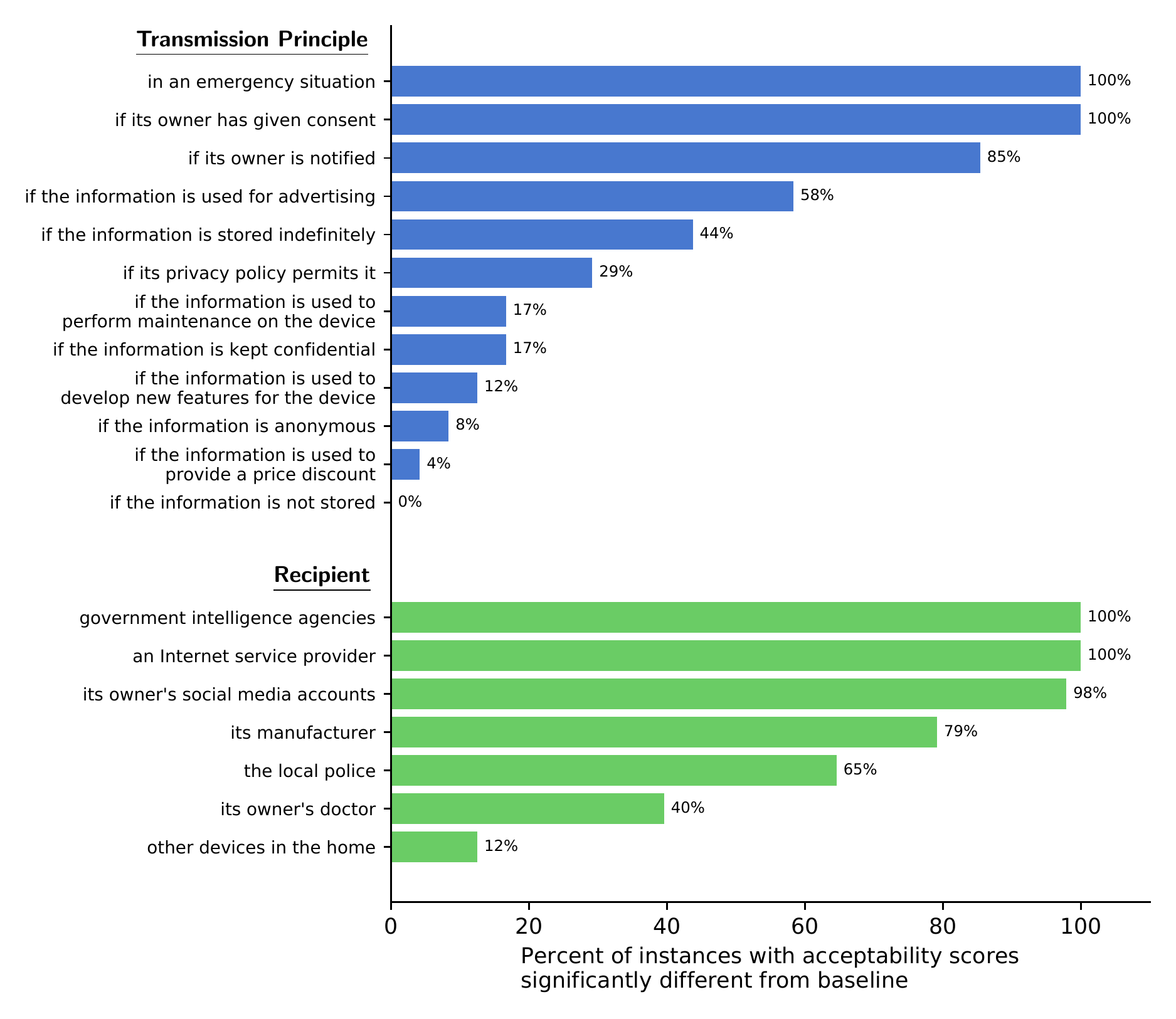}
\caption{
The percentage of instances where the inclusion of the specified transmission principle or recipient resulted in a statistically significant difference in acceptability score compared to the same information flow with the \textit{null} transmission principle or the ``its owner's immediate family'' recipient.}
\label{tbl:sigTP}
\end{figure}

Using the significance test described in Section~\ref{sec:methods-stats}, we evaluate the effect of adding transmission principles to unconditional information flows. We calculate the percentage of instances for which the inclusion of a particular transmission principle (as opposed to the \textit{null} transmission principle) results in a statistically significant difference in average acceptability score (Figure~\ref{tbl:sigTP}).  We also perform the same procedure for recipients, comparing average acceptability scores to information flows with the baseline recipient ``its owner's immediate family" (Figure~\ref{tbl:sigTP}).

The transmission principles ``if its owner has given consent,'' ``in an emergency situation,'' and ``if its owner is notified'' result in significantly different scores 
in $\ge\!85\%$ of instances. This indicates that these parameters heavily influenced participants reactions to information flows. In comparison, the transmission principles ``if the information is not stored,'' ``if the information is used to provide a price discount,'' and ``if the information is anonymous,'' resulted in significantly different scores in $<\!10\%$ of instances. 

The recipients ``government intelligence agencies,'' ``an Internet service provider,'' and ``its owner's social media accounts'' result in significantly different scores
in $\ge\!98\%$ of instances. In comparison, the recipient ``other devices in the home'' resulted in significantly different scores in only $12\%$ of instances. 

\subsection{Effect of Smart Home Device Ownership}
\begin{figure}[tp]
\includegraphics[width=0.9\textwidth]{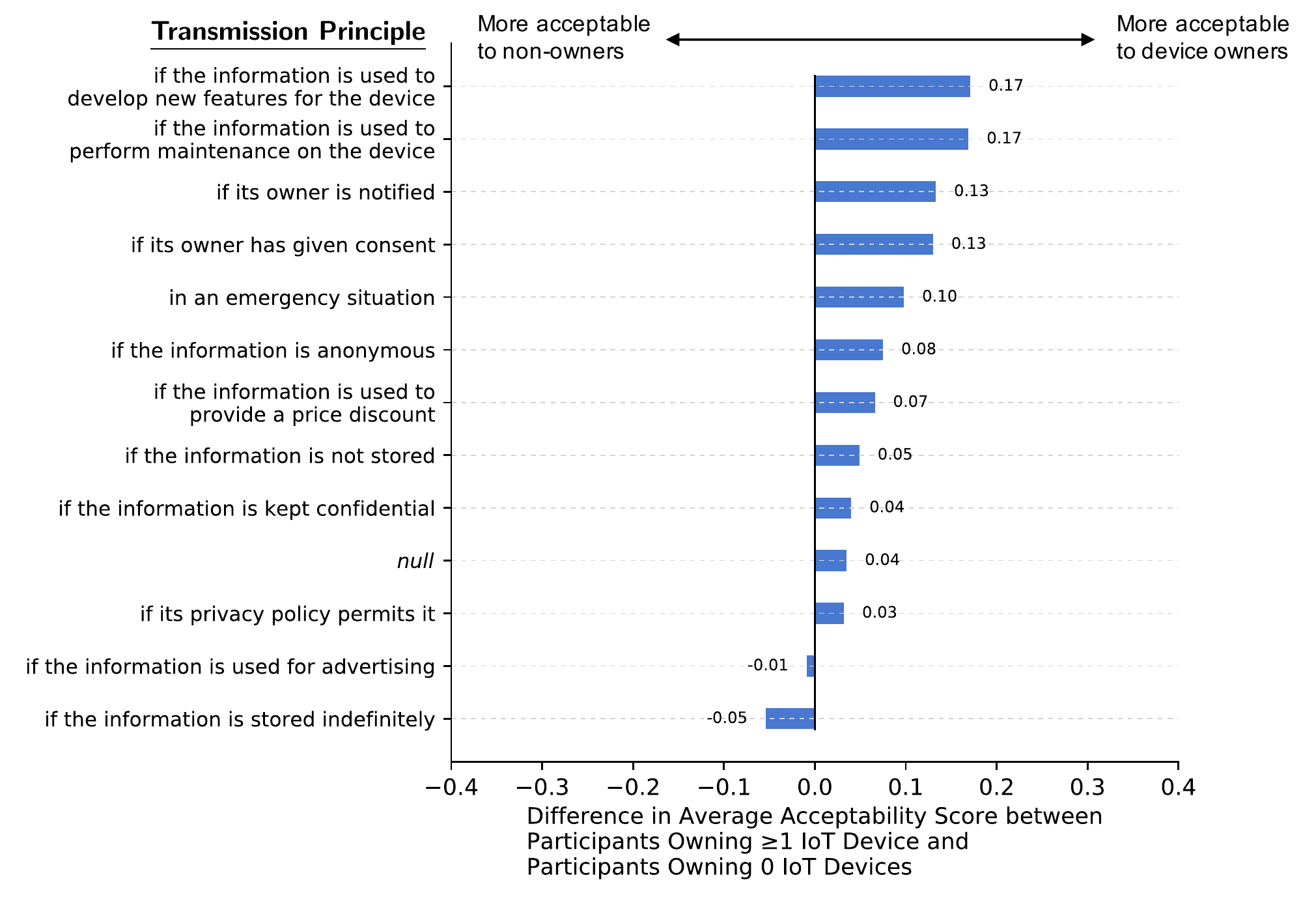}
\caption{Difference in
average acceptability scores of information flows with specified transmission principles between respondents who do and do not own any home IoT devices. 
Positive values indicate that respondents owning IoT devices rated the associated information flows as more acceptable on average.}
\label{tbl:TPdevOwners}
\end{figure}

Figure~\ref{tbl:TPdevOwners} compares the average acceptability scores of information flows with each transmission principle between the 36\% of respondents who self-reported that they own at least 1 smart home device and the 62\% of respondents who own 0 devices. The remaining 2\% of respondents who answered ``I don't know'' to the device ownership question were omitted from this comparison. 

Owning a smart home device generally increased participants' average acceptability scores by a small amount. This effect is most pronounced on information flows with the transmission principles ``if the information is used to develop new features for the device'' ($\Delta0.17$) and ``if the information is used to perform maintenance on the device'' ($\Delta0.17$). The strongest exception is the transmission principle ``if the information is stored indefinitely'' which received higher acceptability scores from participants who do not claim to own any smart home devices ($\Delta0.14$). 
However, there are no transmission principles 
for which owning smart home devices changed the average acceptability score from negative to positive or vice versa. 

\section{Observations and Recommendations}
\label{sec:discussion}
Analysis of our survey responses yields rich insights into IoT privacy norms in the home context. 
The following discussion synthesizes our results into observations and recommendations for IoT device manufacturers, policymakers and regulators. 
These sections, like our choice of information flow parameters, favor breadth over depth in order to demonstrate the range of issues that can be better understood using our survey method. 
We hope that this potential will inspire others to adopt our survey technique and perform their own investigations of contextual privacy norms.

\subsection{Device Manufacturers Should Survey Contextual Privacy Norms}
Suppose a  smart home device manufacturer is designing a product that will collect data about its users and communicate with first and third parties to support a variety of features and revenue sources.
The manufacturer can use its complete knowledge of the device's intended behavior to generate very specific five-parameter information flows involving the device.
The manufacturer can then use our survey method to determine if any of these information flows disrupt established contextual norms. If so, the manufacturer can modify device behavior during the design process before releasing a device that might be subject to consumer backlash.
Even if a manufacturer is unsure what information flows are relevant for their devices, our survey method allows investigation of a broad set of potentially relevant privacy norms for relatively little cost. We were able to survey the acceptability of 3,840 information flows for only \$2,800, a fraction of typical product development budgets. 

Following this recommendation could have prevented several recent privacy-related public relations snafus by IoT device manufacturers. 
For example, knowing that sending information about users' eating habits and home occupancy is especially likely to be deemed unacceptable could have prevented notorious design choices made by  manufacturers of smart TVs and headphones~\cite{matyszczyk2015samsung,2017Bose}.

As a detailed example of how using our survey method could have protected device manufacturers, consider the high-profile cases of IoT toys ``spying'' on children \cite{motherboard-teddy}.  Some of the backlash to these devices is due to their insecure implementations, allowing hackers to obtain sensitive user information. However, many consumers were justifiably upset that these devices collected and indefinitely stored audio recordings of their children regardless of security considerations. The manufacturers of these toys could have used our survey method to determine how to better fit device features to user privacy norms. In this case, surveyed information flows would all have the same sender, attribute, recipient, and subject parameters. The sender would be the toy itself. The recipient would be the device manufacturer. The attribute would be recorded audio, and the subject would be the child playing with the toy. The transmission principles would then vary to reflect different implementation options, such as ``if the information is automatically deleted within 24 hours,'' ``if the information can be manually deleted by an adult user through a web interface,'' etc. 
Because A/B testing of privacy-related designs is difficult to do ethically, such a survey would allow for a data-driven decision about which potential implementation to choose. 

\subsection{Privacy Norms Support Restricting IoT Device Communications}
A recent report on IoT security and privacy by the Broadband Internet Technical Advisory Group (BITAG) recommends that ``IoT devices should be restrictive rather than permissive in communicating'' \cite{bitag}. 
The negative acceptability scores of most information flows in our survey support this recommendation, indicating that restricting IoT device communications is in keeping with 
consumer privacy norms.

This reinforces the importance of critically evaluating device communications during the software development process in order to avoid communications not strictly necessary for core device behavior.
Manufacturers should be especially careful of third-party services and libraries used by their devices, which often invoke their own information flows to potentially unknown recipients. 

\subsection{Privacy Policies Should Clearly State Transmission Principles}
Our results complement another recommendation from the BITAG report: ``IoT devices should ship with a privacy policy that is easy to find \& understand'' \cite{bitag}. 
The use and utility of privacy policies has been examined by many prior research efforts~\cite{martin2015privacy, cranor2000beyond,jensen2004privacy,earp2005examining,egelman2009timing}. 
Previous studies have shown that ``consumers believe the term `privacy policy' on a website means that the site protects their privacy''~\cite{staff2010protecting}. 
In our results, information flows with the transmission principle ``if its privacy policy permits it''
are more acceptable than unconditional flows 
but still had negative average acceptability scores across eight out of nine recipients. 
Simply abiding by a privacy policy does not seem to make an IoT device automatically adhere to consumer privacy norms, regardless of the policy's clarity. 
However, easily accessible and interpretable privacy policies could allow manufacturers to benefit from other positive effects on information flow acceptability observed by our survey. For example, information flows that occur only in emergency situations were generally deemed as acceptable, while their unconditional counterparts were generally unacceptable. If a device communicates certain information only in emergency situations, its manufacturer should use the privacy policy to clearly highlight this criteria to reduce consumer privacy concerns.
This is particularly relevant for IoT security systems, such as smart smoke detectors, carbon monoxide detectors, and glass break detectors, as well as for certain health monitoring devices, such as fall detectors for the elderly.

\subsection{User Consent is Broadly Important for Information Flow Acceptability}
The transmission principle, ``if the owner has given consent'' has the highest average acceptability scores across all recipients and is the only information flow parameter with positive average acceptability scores across all conditions. 
In other words, survey respondents tended to accept information transfers with user consent irrespective of the rest of the context. 
This result aligns 
with notice-and-consent initiatives advocating for meaningful and timely data usage notifications and for requiring user consent to collect information~\cite{FIPP}. 

However, our results also provide a clear illustration of the idea argued by Nissenbaum in~\cite{nissenbaum2010privacy} that consent-based definitions of privacy are just a part of a bigger picture. For example, information flows taking place ``in an emergency situation'' had positive acceptability scores with five of six recipients without any specification of consent. 
Such insights further support calls to adopt new a privacy paradigm for IoT that will ``respect the context in which personally identifiable information is collected''~\cite{newIoTParadigm2013}.

\subsection{Local Data Sharing Should Consider Secondary Information Flows}
The recipient ``other devices in the home'' was the second most acceptable recipient in the survey and the only recipient with a positive average acceptability score in information flows with the \textit{null} transmission principle.
This may indicate there is a broader privacy norm distinguishing recipients inside the home versus outside the home. This is promising  for manufacturers of interconnected smart home IoT systems that share information among a network of devices.

However, device manufacturers should also consider secondary information flows from their devices through an intermediary.  At first glance, the increased acceptability of information flows to other devices in the home might indicate that devices that only communicate with an IoT hub or smartphone by Bluetooth or local WiFi can be less concerned about sending user information. However, a device that sends its owner's location to a nearby smartphone and then the smartphone sends the location to a manufacturer's cloud server would have participated in two information flows: a flow with the recipient ``other devices in the home'' and a flow with the recipient ``its manufacturer.'' While the first flow may be acceptable given user privacy norms, the second may not. This is particularly relevant for manufacturers of single-purpose sensors and actuators designed to connect to an IoT hub made by a different company. Such manufacturers should avoid having their devices send potentially sensitive information with greater frequency or precision than necessary. Obfuscation techniques could also be employed to prevent privacy violating information flows from an intermediary device.

\subsection{User Familiarity and Market Penetration Impact Privacy Norms}
The power meter, fitness tracker, and thermostat 
have the least negative acceptability scores averaged across all tested attributes, while the refrigerator, security camera, and door lock have the most negative.
We may speculate that these results reflect exposure to smart devices of particular types.  
Fitness trackers and Nest-like smart thermostats, widely available and well-advertised, have adjusted users' expectations with regards 
to the connectivity and information sharing behavior of these devices. 
Smart refrigerators and door locks have seen comparatively less market penetration.  
This interpretation suggests that greater exposure to specific types of smart home devices has modified or created privacy norms, making information flows involving these devices more acceptable. 
However, it would require 
further longitudinal surveys to indicate whether privacy norms are indeed being modified by exposure to IoT devices over time.  
Such surveys, especially focusing on a single sender with a wider variety of other information flow parameters, would be valuable for IoT device manufacturers to see how their products fit into evolving cultural norms. This could give an indication of when or how gradually to release new products or new features.  

\subsection{Device Communications Should Directly Support Primary Functionality}
Examining acceptability scores across pairs of senders and attributes indicates that information flows may be more acceptable 
when the attribute is more closely related to the primary function of the device.
For example, a fitness tracker is likely to need heart rate data and exercise routine information in order to provide fitness recommendations. These information flows are correspondingly more acceptable than a fitness tracker sending less fitness-relevant information, such as recorded audio or home occupancy.
Similarly, a smart refrigerator sending users' eating habits (associated with the food-related purpose of the device) is deemed more appropriate than a refrigerator sending users' locations. 
However, the effect size is small and not consistent across all sender/attribute pairs.
These results motivate more extensive follow-up surveys to disentangle pairwise sender/attribute effects on privacy norms. 

\subsection{ISP Data Collection, Advertising, and Indefinite Data Storage Generally Violate Privacy Norms}
ISPs 
are among the least acceptable recipients across all transmission principles. 
A comparison of the median value of the average acceptability score for the ISP recipient against the the median value of the average score for the baseline recipient ``its owner's immediate family'' ($-2$ vs. $0$) further indicates that the ISP recipient has a negative effect on information flow acceptability. 
This result is relevant given the recent nullification  of the FCC broadband consumer privacy resolution~\cite{fccResolution}, which effectively allows ISPs to mine and sell customer information. Our results provide a strong indication that consumers do not favor the idea of devices transmitting information to ISPs despite this regulatory action.

The transmission principles ``if the information is used for advertising'' and ``if the information is stored indefinitely'' had the lowest acceptability ratings averaged across all recipients and were the only transmission principles to have lower average scores than unconditional flows with the \textit{null} transmission principle. 
User dislike of sharing data for advertising aligns with previous studies~\cite{phelps2000privacy}, and a mistrust of persistent cloud storage reflects recent high-profile data breaches.  However, 
advertising and indefinite  storage (for a variety of purposes) are still standard for many companies collecting data from non-IoT provenances. Our  survey indicates that these privacy concerns, originally developed in non-IoT contexts, are still very relevant to the nascent home IoT. 
This recommends that IoT manufacturers remain cautious with their data storage and advertisement practices or risk breaching privacy norms and alienating users.

\subsection{IoT Device Owners Are More Accepting of Smart Home Information Flows}
Responses to the technical background questions allow us to compare privacy expectations between participants who own one or more IoT devices in their homes compared to those who do not. 
We chose to compare responses across transmission principles, because all participants answered questions about information flows with all transmission principles. 

Acceptability scores are higher for participants owning smart home devices across all transmission principles, but the differences between respondent groups are small.
These differences motivate follow-up studies to see whether this effect is consistent across larger populations. These studies could be combined with standardized privacy concern scales, such as IUIPC \cite{malhotra2004internet}, to see how people with different underlying privacy concerns and different exposures to novel technologies react to information flows. 
More fundamentally, this line of research would 
follow the original formulation of CI, which calls for examining established norms in terms of their ``merits as a function of their meaning and significance in relation to the aims, purposes, and values of the context''~\cite{nissenbaum2010privacy}.

\subsection{Smart Home Privacy Norms Exhibit Similarities to Smartphone Privacy}
Our results also warrant a comparison with existing research that has explored various dimensions of mobile application permissions \cite{linEandP,linRestoring,primalCI,pewsmartphone}. Previous studies \cite{linRestoring,linEandP} have shown that smartphone users are generally uncomfortable sharing their data---including their location, contacts, and account data---with advertisers and social networks services. Users also cite concerns about personal data as reasons to avoid downloading applications \cite{pewsmartphone}. The vast majority of smartphone participants have been shown to deny at least one information flow---as defined by CI---in current smartphone systems \cite{primalCI}.

While only some of the information flows that we considered directly map onto current flows in smartphone systems, we noted some similarities. For example, our participants found  flows with the transmission principles ``if the information is used for advertising''  and ``its owner's social media accounts'' largely unacceptable~\cite{linEandP, linRestoring}. 
Nevertheless, our participants were generally comfortable with flows that were conditioned on consent.  
Future studies could use our survey methodology with ``smartphone'' as a sender to enable a more direct comparisons between smartphone privacy norms and those we discovered for IoT devices.
Combining such surveys with field experiments and interventions~\cite{primalCI} could also indicate how and when smartphone users will deny consent to specific flows.

\section{Limitations and Future Work}
\label{sec:future}
We acknowledge a number of limitations of our approach, which we believe can be addressed with future studies and refinement of CI-based survey methods.

\subsection{Parameter Ambiguity}
Our goal was to demonstrate that the CI survey method is effective in broad contexts, even if some interpretation of information flow parameters is left to survey participants. Such broad contexts are likely of interest to regulators or consumer advocates seeking holistic overviews of privacy norms.

Nevertheless, we recognize that participant interpretations may have caused uncontrolled variations in information flow acceptability scores.
For example, ``its owner's immediate family'' could be interpreted as various nuclear to extended family groups. 
The generic device types (e.g., ``thermostat'' rather than ``Nest Learning Thermostat'') may have caused participants to envision different on-market devices with different features or to imagine entirely non-existent devices.
However, the acceptability scores of information flows that did include a specific name-brand device (``a personal assistant (e.g. Amazon Echo)'') had a standard deviation within 0.02 points of the standard deviations of flows with all other senders. This suggests that using generic device types did not significantly increase acceptability score variability, but follow-up surveys with overlapping device types and specific devices 
would be needed to confirm.

Device manufacturers or regulatory investigators could use our CI survey method with more clearly defined information flow parameters to discover expectations about product behavior in specific scenarios. We look forward to seeing such applications of our survey method in future research. 

\begin{table}[tp]
\footnotesize
\caption{Selected representative open-ended comments relevant to participant rationale categorized by general value of concern. Participant ID numbers in parentheses.}
\begin{tabular}{p{0.11\textwidth}|p{0.79\textwidth}}
\textbf{\normalsize Value}\vspace{2pt} & \textbf{\normalsize Participant Comments}\vspace{2pt}\\
\hline
\textbf{Trust} & \vspace{-7pt}\begin{itemize}[leftmargin=8pt]
\item  I'm starting not to trust smart technology because of the capability that ``big brother'' may be listening. (P513)
\item I don't trust my ISP or my government or companies.  I used to trust my last doctor but not my current one. (P863)
\end{itemize}\vspace{-10pt}\\
\hline
\textbf{Security} & \vspace{-7pt}\begin{itemize}[leftmargin=8pt] 
\item  I do not and will not own smart appliances due to the danger of hacking. Sending info in any situation is absolutely completely unacceptable to me. (P302)
\item I am personally leery about connected devices that keep information, especially considering their lack of security. (P1623)
\end{itemize}\vspace{-10pt}\\
\hline
\textbf{Privacy} & \vspace{-7pt}\begin{itemize}[leftmargin=8pt] 
\item  I personally value my privacy more than convenience given the option. (P407)
\item  For me it's totally unacceptable, for any reason, to have audio on my thermostat. A person should be speak in his/her own home without being recorded by anything or anybody. (P1167)
\item ISPs should never have access to anyone's physical location. It is idiotic to alert social media where you are, especially if you are traveling. (P1494)
\item I understand that ``privacy'' is really an oxymoron but I'm sad (and a little concerned) that we're not even safe in our own homes from information gathering devices. (P1253)
\end{itemize}\vspace{-10pt}\\
\hline
\textbf{Transparency} &\vspace{-7pt}\begin{itemize}[leftmargin=8pt]
\item Sharing personal information is only ever completely acceptable when the owner gives consent. (P1749)
\item Even if consent is given, I still have privacy concerns about some of those situations. (P983)
\item In general I think transmitting that information is unacceptable, although owner consent \& emergencies may be okay if that info is transmitted to police, doctors and immediate family. (P1665)
\item Just a thought I had: a customer can give consent by checking that they agree to privacy policy, while not having read the policy. On reflecting on this, I think I may be more careful to agree that it's okay if a person gives consent this way. (P1282)
\item Storing and sharing any of the information in any of the situations is unacceptable unless the individual consumer for some reason wants them to, is clearly informed what they're agreeing to, and gives clearly expressed permission. (P79)
\item I believe many people will volunteer to provide their data and that is perfectly okay and the way we should move towards. (P917) 
\end{itemize}\vspace{-10pt}\\
\end{tabular}
\label{fig:open-ended-response}
\end{table}

\subsection{Participant Rationale}
Our CI survey method offers a way to discover privacy norms regarding information flows. An even deeper understanding would require evaluating  participant rationale and the tradeoffs between values (e.g., trust, national security, safety and/or security) and aims/purposes (e.g., making a system more efficient or a user more productive) that contribute to the emergence of the norms.

We examined participants' optional open-ended comments for indications of the values motivating their responses. About 30 participants gave comments related to the IoT content of the survey. Several of these express general concerns about privacy (e.g., ``Worried about this type of technology in the future''). Those that do provide further insight are collected in Table~\ref{fig:open-ended-response}. Many say that information sharing from IoT devices is either never acceptable or acceptable only with owner consent or in emergencies. This matches the aggregated acceptability scores from the full survey results. Concerns about IoT device security and lack of trust in various entities (ISP, government, companies, smart technology, doctor) are cited as reasons for disapproving information flows. 
Ultimately however, the limited number and length of these responses restricts their ability to explain nuanced variations in acceptability scores across information flows.
Researchers who use our CI survey method to systematically appraise a wide space of privacy norms may elect to perform follow-up studies to identify the values underlying particularly interesting or surprising discovered norms.  

\subsection{Complex Modeling} 
Users' privacy expectations likely depend on additional factors beyond the five CI information flow parameters (e.g., demographics information). The regular format of responses from our survey methodology makes it possible construct statistical models to test the interactions of these factors. However, the complexity of such models increases substantially with each new factor. With sufficient data and more complex models (e.g., mixed-effect models~\cite{bates2014fitting}), future research could identify how other factors influence privacy norms.

\subsection{Limitations of the Platform} 
Because we conducted our survey on Amazon Mechanical Turk, our results may be less generalizable to the broader US population. However, while the MTurk population is limited in its diversity, research \cite{bartneck2015comparing,simons2012common} has shown that it is similar to participants from university campuses and other online participant pools. Future research could validate and extend our findings to more diverse participants.

\subsection{Online Survey Tool}
Our next goal is to create an online tool which will allow anyone to easily create and conduct surveys using our CI method. Users will be able to input CI parameters relevant to their particular domain, choose from provided default parameters, and specify parameter combinations to exclude. The tool will automatically generate an online survey using our format and deploy the survey on MTurk. We will provide an interface that plots results, highlights notably acceptable or unacceptable parameters or information flows, and allows the user to re-deploy the survey at a later date to collect longitudinal data. We hope that the results of this research, combined with a user-friendly online tool, will incite further studies using our CI survey method. 

\section{Related Work}
\label{sec:related}
Previous work has captured and analyzed privacy norms in the IoT context. This section surveys the work that is most closely related to this study.

In 2012, Barkhuus highlighted the shortcomings of the existing privacy frameworks on self-gathered empirical data~\cite{barkhuus2012mismeasurement}. She argued for the use of CI for privacy-related user studies as a way to understand the ``contextually grounded reasons for people's privacy concern or lack thereof.'' This observation motivates our work.

In the same year, Winter performed a small study using CI to identify specific practices in an IoT setting that could be perceived as privacy violations~\cite{winter2012privacy}. Our work also uses the CI framework to construct context-related questions that can help identify practices that conflict with established privacy norms. However, our study examines many more actors, attributes, and transmission principles. We also offer a more thorough statistical analysis of responses.

A 2013 study conducted by Pew Research Center, which surveyed 461 United States (US) adults and facilitated 9 online focus groups with 80 people each, found that Americans would opt to ``share personal information or permit surveillance in return for getting something of perceived value''~\cite{prc_survey}. A section on ``home activities, comfort and data capture'' specifically explored users' perceptions towards a smart thermostat that can ``learn about your temperature zone and movements around the house and potentially save you on your energy bill\ldots\!in return for sharing data about some of the basic activities that take place in your house.'' Data analysis showed that 55\% of participants found this behavior unacceptable for various reasons, while 27\% deemed it acceptable. Our survey provides a much more comprehensive analysis of such scenarios and covers a range of settings, devices, and information types to construct a more complete picture of the privacy norms surrounding IoT information flows. 

In 2015, Horne et al.\! studied emerging privacy norms and user privacy expectations in response to new technology~\cite{horne2015privacy}. In three separate vignette experiments, participants were recruited on MTurk to answer a series of questions about the frequency and granularity of information collection by a smart power meter. The questions asked whether the confidentiality or sale of the collected information affected their privacy expectations. Our work also explores privacy expectations and norms, but we rely on the CI framework to rigorously scale our data collection to many more users and settings. Our method allows us to ask and examine users' privacy expectations for different devices, information recipients, and conditions under which information is shared. 

A 2016 study by Martin and Nissenbaum~\cite{martin2016measuring} surveyed 569 respondents using a series of vignettes describing different CI information flows with varying receivers and transmission principles. This study found that specifying additional contextual information affects users' perception of what information is sensitive. They conclude that knowing ``how the information is used is more important to meeting/violating privacy expectations than the type and sensitivity level of given information.'' While our work also uses CI to provide additional contextual information to capture users' privacy expectations, it offers a robust methodology that 
uses all the CI parameters.

Our survey method is based on previous work from 2016 by Shvartzshnaider et al. that used the language of CI to crowdsource discovery of users' privacy expectations in the education domain~\cite{shvartzshnaider2016learning}. 
Users were asked multiple yes-or-no questions such as
``Is it acceptable for the student's professor to share the student's transcript with the student's academic advisor if the student is performing poorly?''
The aggregated responses were analyzed to produce a set of most approved and most disapproved norms in the surveyed community. 

In 2017, Emami-Naeini et al.\! used a 1,007-participant vignette study to capture privacy expectations and preferences of users in a set of 380 IoT use-case scenarios~\cite{soupsIoTprivacy}. MTurk workers were presented with 14 different IoT data collection scenarios, including factors such as location, biometrics, temperature, purpose, retention time, and whether the data is shared after collection. A linear model provides an indication of which factors positively or negatively affect users' comfort levels.  
Overall, this work is timely and shares a similar motivation to ours, but does not use a formal theory of privacy and tests many fewer conditions than the 3,840 information flows surveyed in our study.
By showing that "privacy preferences are diverse and context-dependent,'' this work supports our use of CI.

\section{Conclusion}
\label{sec:conclusion}
This work presents a survey method for privacy norm discovery that integrates a formal theory of privacy with combinatorial testing at scale.
We use the method to discover privacy norms regarding smart home IoT devices. Our survey of 1731 U.S.\! adults and 3,840 information flows provides actionable recommendations for device manufacturers, regulators, and consumer advocates.
These results demonstrate that our CI survey method enables effective discovery of privacy norms, even in rapidly-changing technological domains. 
Our method is easily adaptable to arbitrary contexts with varying actors, information types, and communication conditions, paving the way for future studies informing the design of emerging technologies.

\begin{acks}
We thank Helen Nissenbaum and Marshini Chetty.
This work is supported by the Department of Defense through the National Defense Science and Engineering Graduate Fellowship (NDSEG) Program, a Google Faculty Research Award, the National Science Foundation through awards CNS-1535796 and CNS-1539902, and the Princeton University Center for Information Technology Policy Internet of Things Consortium.
\end{acks}

\bibliographystyle{ACM-Reference-Format}
\bibliography{bibliography}

\begin{table}[h]
\section*{Appendix}
\vspace{12pt}
\footnotesize
\caption{Self-reported demographics and technical background of survey participants. }
\begin{tabular}{rl|rl}
\textbf{Metric} & \textbf{Sample} & \textbf{Metric} & \textbf{Sample} \\
\toprule
Female     & 51\%  & Live with family  & 61\% \\
Male       & 49\%  & Live alone        & 20\% \\
Other      & $<$1\% & Live with one or more non-family roommates    & 16\%\\
Prefer not to disclose   & 1\%  & Other  & 2\% \\
& & Prefer not to disclose                          & 1\% \\

Democrat & 44\% & &\\ 
Republican  & 21\% & A one-family house detached from any other house  & 60\% \\
Independent & 33\% & A building with 10 or more apartments & 14\% \\
Prefer not to disclose & 3\% & A building with fewer than 10 apartments & 11\%\\
& & A one-family house attached to one or more houses   & 8\% \\

Less than \$10k      & 4\% & A mobile home & 3\% \\
\$10k - \$20k        & 8\%  & A dormitory  & $<$1\% \\
\$20k - \$30k        & 12\% & A boat, RV, van, etc.  & $<$1\% \\
\$30k - \$40k        & 11\%  & Other   & 1\% \\ 
\$40k - \$50k        & 11\% & Prefer not to disclose  & 1\% \\
\$50k - \$60k        & 11\% & \\
\$60k - \$70k        & 10\% & Suburban    & 53\% \\
\$70k - \$80k        & 8\% & Urban       & 30\% \\
\$80k - \$90k        & 5\% & Rural       & 18\% \\
\$90k - \$100k        & 5\% & \\ 
\$100k - \$150k      & 9\% & Married or domestic partnership    & 46\% \\
More than \$150,000        & 3\% & Single, never married  & 44\% \\
Prefer not to disclose     & 2\% & Divorced     & 7\% \\
& & Separated    & 1\% \\

18-24 years old           & 10\% & Widowed          & 1\% \\
25-34 years old           & 45\% & Prefer not to disclose  & 1\% \\
35-44 years old           & 23\% & \\
45-54 years old           & 11\% & No children under 16     & 67\% \\
55-64 years old           & 6\% & Children under 16        & 32\% \\
65-74 years old           & 2\% & Prefer not to disclose   & 1\% \\
75 years or older         & $<$1\% &\\
Prefer not to disclose    & 1\% & 0-3 hours Internet use per day & 15\% \\
&& 4-7 hours Internet use per day   & 45\% \\

Nursery school to 8th grade  & $<$1\% & 8-12 hours Internet use per day   & 29\% \\
Some high school, no diploma  & 1\% & $>$12 hours Internet use per day  & 11\% \\
High school graduate, diploma or the equivalent   & 9\% & \\
Trade/technical/vocational training & 3\% & Own 0 IoT devices  & 62\% \\
Some college credit, no degree & 23\% & Own $\ge$1  IoT device & 36\% \\
Bachelor's degree  & 40\% & I don't know  & 2\% \\
Associate degree   & 12\% & \\
Master's degree    & 10\% & I set up my IoT devices     & 78\% \\
Professional degree  & 2\% & Someone else set up my IoT devices  & 21\% \\
Doctorate degree    & 1\%  & I don't remember who set up my IoT devices  & 1\% \\
Prefer not to disclose                            & 1\% \\
&\\

\end{tabular}
\label{fig:demographics}
\end{table}

\end{document}